\documentclass[12pt]{myels}
\usepackage{epsfig}
\newcommand{\nn}{\nonumber}
\newcommand{\raw}{\rightarrow}
\newcommand{\lraw}{\leftrightarrow}

\newcommand{\be}{\begin{equation}}
\newcommand{\ee}{\end{equation}}
\newcommand{\bea}{\begin{eqnarray}}
\newcommand{\eea}{\end{eqnarray}}

\newcommand{\evb}{ {\rm eV$^2$} }

\newcommand{\PCPV}{
\begin{picture}(22,10)
\put(8,-2){\line(2,1){12}}
\put(0,0){$P_{CP}$}
\end{picture}}
\newcommand{\PCPC}{
\begin{picture}(22,10)
\put(0,0){$P_{CP}$}
\end{picture}}
\begin{document}



\begin{frontmatter}
\begin{center}

\title{Four species neutrino oscillations at $\nu$-Factory: 
       sensitivity and CP-violation}




\begin{center}
\author[a]{A. \snm Donini}\footnote{E-mail donini@daniel.ft.uam.es}, 
\author[a]{M.B. \snm Gavela}\footnote{E-mail gavela@garuda.ft.uam.es}, 
\author[b]{P. \snm Hern\'andez}\footnote{On leave from Departamento de 
F\'{\i}sica Te\'orica, Universidad de Valencia. 
E-mail Pilar.Hernandez@cern.ch} and 
\author[a]{S. \snm Rigolin}\footnote{E-mail rigolin@daniel.ft.uam.es}
\address[a]{Departamento de F\'{\i}sica Te\'orica C-XI, Facultad de Ciencias, 
Universidad Aut\'onoma de Madrid, \cty Cantoblanco, Madrid 28049, \cny Spain}
\address[b]{CERN, \cty 1211 Geneve 23, \cny Switzerland}
\end{center}



\begin{abstract}

The prospects of measuring the leptonic angles and CP-odd phases at a 
{\em neutrino factory} are discussed in the scenario of three active plus 
one sterile neutrino. We consider the $\nu_\mu \raw \nu_e$ LSND signal.
Its associated large mass difference leads to observable
 neutrino 
oscillations at short ($\sim 1$ km) baseline experiments. Sensitivities 
to the leptonic angles down to $10^{-3}$ can be easily achieved with a 
1 Ton detector. Longer baseline experiments ($\sim 100$ km) with a 1 Kton 
detector can provide very clean tests of CP-violation especially 
through tau lepton detection.  
%
%
\end{abstract}



\end{center}
\end{frontmatter}


\setcounter{footnote}{0}

%
\section{Introduction}
%
%
Indications in favour of neutrino oscillations have been 
obtained in both the atmospheric and solar neutrino experiments 
\cite{Superka,sun,sunSK,atm}. The atmospheric neutrino data require 
$\Delta m_{atm}^2 \sim (0.3 - 7) \times 10^{-3}$ \evb whereas the solar 
neutrino data prefer either $\Delta m_{sun}^2 \sim 10^{-5} - 10^{-4}$ 
\evb if interpreted as MSW (matter enhanced) oscillations,
 or $\sim 10^{-10}$ \evb 
if the vacuum solution is preferred by nature. The LSND data \cite{LSND}
 would indicate a $\nu_\mu \to \nu_e$ oscillation with a 
third, very distinct, neutrino mass difference: $\Delta m_{LSND}^2 \sim 
0.3 - 1$ \evb. 
The usual three neutrino picture 
has then to be enlarged to explain the present ensemble of data: four different 
light neutrino species are needed. The new light neutrino is usually 
denoted as sterile \cite{sterile,sterosc}, in it cannot 
have the standard weak interaction charge in order to comply with 
the strong bounds on the $Z^0$ invisible decay width \cite{LEPnu}. 

We derive here convenient parametrizations of the physical 
mixing angles and CP phases in the case of four neutrino species, 
and study their experimental signals at the {\em neutrino factory}\footnote{
For a detailed analysis of the three family scenario see 
\cite{dgh,noi,Donini:1999jc,romanino,romanino2}.}. 

For the sake of illustration, we shall consider as a ``reference 
set-up'' the neutrino beams resulting from the decay of $n_\mu = 2 \times
10^{20} \mu^+$s and/or $\mu^-$s in a straight section of an $E_\mu = 
10\,-\,50$ GeV muon accumulator ring. Muon energies of $40\,-\,50$ GeV are at 
present under discussion \cite{lyon} as a convenient goal, as they allow 
good background rejection \cite{jj} and do not preclude the 
exploration of neutrino signals at lower energies. This is because
the number of neutrinos in a given energy bin does not depend on the 
parent muon energy, while the total number of oscillated and interacted
neutrinos increases with $E_\mu$ \cite{muring}. As the dominant 
signals are expected to peak at ``LSND'' distances, most of the parameter
space can be explored in experiments at short baseline (SBL) distances of 
O(1) km and with a very small 1 Ton detector. To optimize the observability 
of CP-violation asymmetries larger distances (O(10-100) km) and larger 
detector (1 kTon) are needed.

In section 2 we describe our parametrization.
Sections 3 and 4 discuss the experimental sensitivities to angles and
CP phases, respectively.
We conclude in section 5.

%
\section{Four-Neutrino species}
\label{sect:fourneutrino}
\begin{figure}[t]
\begin{tabular}{lcr}
\epsfig{file=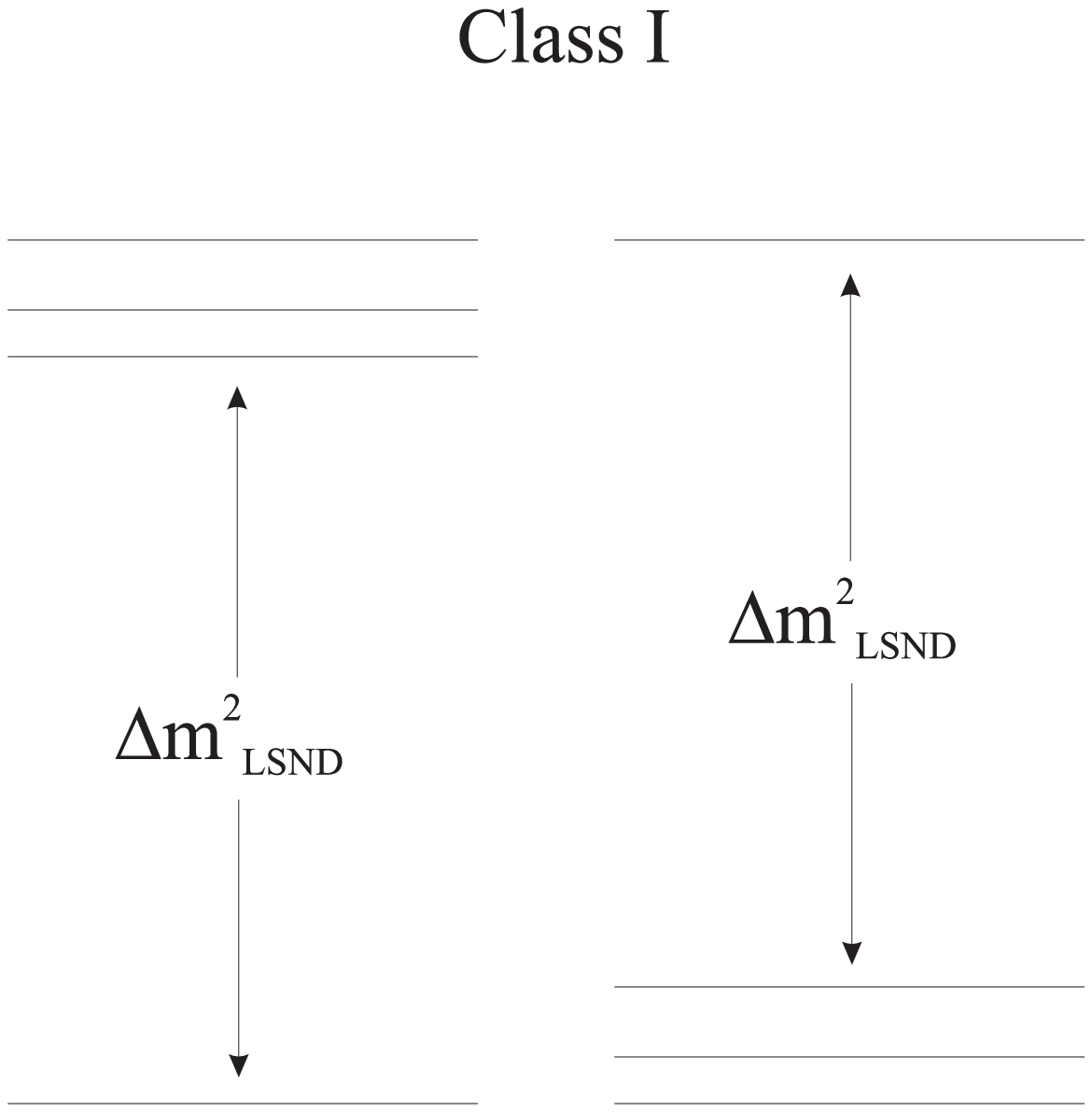, width=6.7cm} & \mbox{ \hskip 0.7cm} &
\epsfig{file=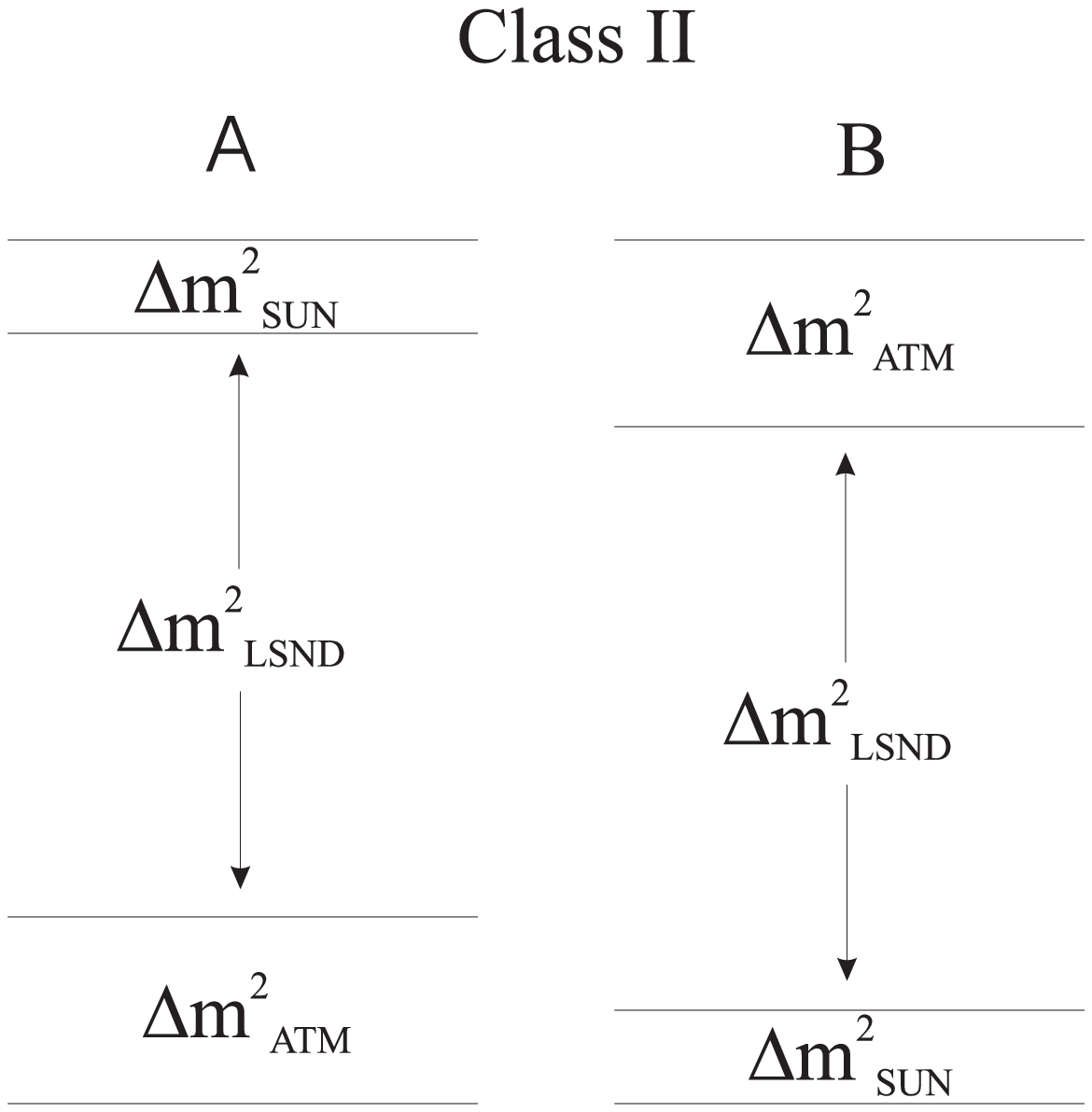, width=6.7cm} 
\end{tabular}
\caption{{\it Different type of four-neutrino families scenarios: 
Class-I scenarios (left); Class-II scenarios (right).}} 
\label{fig:4famtyp}
\end{figure}

%
%
The masses of the four neutrino species 
can be ordered in two ways, as depicted in 
Fig.\ref{fig:4famtyp}:
\begin{enumerate}
\item 
Three almost degenerate neutrinos, accounting for solar and atmospheric 
oscillations, separated from the fourth neutrino specie by the LSND mass 
difference;
\item
Two almost degenerate neutrino pairs, accounting each for solar 
and atmospheric oscillations, separated by the LSND mass gap. 
\end{enumerate}
We refer to these possibilities respectively as class-I and class-II schemes. 
Both of them contain a number of sub-classes accounting for all the possible 
ordering of the neutrino species. It has been shown in \cite{bilenky} 
that the combined analysis of solar, atmospheric and LSND experiments tends 
to exclude all class-I schemes. We consider then two possible 
class-II schemes and choose for 
definiteness the hierarchycal class-IIB scenario in Fig.\ref{fig:4famtyp}.
 The lighter pair (1-2) 
is separated by the solar mass difference and the heavier pair (3-4) 
by the atmospheric one\footnote{The alternative option, with a small 
separation in the heavier pair and a larger one 
for the lighter pair, amounts to reversing the sign of the LSND 
mass difference. The analysis below does not depend on this 
choice.}. We work in the convention:
\be
\nu_s \simeq \nu_1 \lraw \nu_e \simeq \nu_2 \qquad 
\Delta m^2_{12} = \Delta m^2_{sol} 
\ee
with the solar deficit assigned to $\nu_s-\nu_e$ oscillations, and
\be
\nu_\mu \simeq \nu_3 \lraw \nu_\tau \simeq \nu_4 \qquad 
\Delta m^2_{34} = \Delta m^2_{atm}
\ee
with the atmospheric anomaly due to $\nu_\mu-\nu_\tau$ oscillations. 
The alternative possibility of identifying the atmospheric anomaly as a 
$\nu_\mu-\nu_s$ oscillation is actually disfavoured by SK data\cite{Superka}.


Whatever the mechanism responsible for neutrino masses,
given $n$ light neutrino species oscillation experiments are 
only sensitive to a unitary $n \times n$ mixing matrix ``a la CKM''.
In all generality, the parameter space of a 
four-species scenario would consist of six rotation angles and three 
complex phases if the neutrinos 
are Dirac fermions, while it spans six angles plus six phases if the 
neutrinos are Majorana fermions. Among the latter, three are pure 
Majorana phases and thus to be disregarded in what follows as they
are unobservable in oscillations, reducing the 
analysis to the mentioned $4 \times 4$ ``Dirac-type'' system. 
 
Within our conventions and the large mass hierarchy indicated by data:
\be
\Delta m_{sol}^2 = \Delta m_{12}^2 \ll 
\Delta m_{atm}^2 = \Delta m_{34}^2 \ll 
\Delta m_{LSND}^2 = \Delta m_{23}^2 ,    
\label{hier4fam}
\ee 
two approximations are useful:
\begin{enumerate}
\item
$\Delta m_{12}^2 = \Delta m_{34}^2 = 0$, ``one mass scale dominance'' (or 
minimal) scheme,   
\item
$\Delta m_{12}^2 = 0$, ``two mass scale dominance'' (or next-to-minimal) 
scheme.  
\end{enumerate}
The number of independent angles and phases is then reduced as reported in 
Tab.~\ref{tab:4fampar}. The minimal scheme is sufficient to illustrate 
the sensitivity to the mixing angles other than the solar and atmospheric 
ones.
The next-to-minimal scheme is necessary to address the question of 
CP-violation, as two non-zero mass differences are required to 
produce observable effects, alike to the standard three-family scenario. 

%
%
\begin{table}[t]
\centering
\begin{tabular}{||c|c|c|c||}
\hline\hline
         & Angles & Dirac CP-phases & Majorana CP-phases \\
\hline\hline
Majorana $\nu$'s &   6   &   3   &  3   \\
\hline 
Dirac $\nu$'s &   6   &   3   &  0   \\
\hline
Dirac $\nu$'s   &   5   &   2   &  0   \\
$\Delta m_{12}^2 = 0$ & & & \\
\hline
Dirac $\nu$'s   &   4   &   1   &  0   \\
$\Delta m_{12}^2 = \Delta m_{34}^2 = 0$ & & & \\
\hline \hline
\end{tabular}
\caption{\it{Parameter space for four neutrino families: for Dirac
neutrinos we consider the general case with three non-zero mass differences
and the particular case (considered in the rest of the paper) with
one or two mass differences set to zero; for Majorana neutrinos we
consider only the general case.}}
\label{tab:4fampar}
\end{table}
%
%

A general rotation in a four dimensional space can be obtained by performing
six different rotations $U_{ij}$ in the $(i,j)$ plane, resulting in plenty 
of different possible parametrizations of the mixing matrix, disregarding 
phases. We choose the following convenient parametrization, given the 
hierarchy of mass differences in eq. (\ref{hier4fam}):
\be
U = U_{14} (\theta_{14}) U_{13} (\theta_{13}) U_{24} (\theta_{24}) 
    U_{23} (\theta_{23},\delta_3) U_{34} (\theta_{34}\delta_2) 
    U_{12} (\theta_{12},\delta_1).
\label{ourpar}
\ee
As shown in Table 2, if a given mass difference vanishes the number of
physical angles and phases gets reduced by one. A convenient parametrization 
of the angles is that in which the rotation matrices corresponding to the 
most degenerate pairs of eigenstates are placed to the extreme right. If the 
eigenstates $i$ and $j$ are degenerate and the matrix $U_{ij}$ is located
to the right in eq.~(\ref{ourpar}), the angle $\theta_{ij}$ becomes 
automatically unphysical. When a different ordering is 
taken no angle disappears from the oscillation probabilities. A redefinition of
the rest of the parameters would then be necessary in order to illustrate the 
remaining reduced parameter space in a transparent way.  
Our parametrization corresponds thus to the ``cleanest'' choice, having settled 
at the extreme right the rotations corresponding to the most degenerate pairs. 

In the ``one mass dominance'' scheme,  the pairs $(\theta_{12}, \delta_1)$ and 
$(\theta_{34}, \delta_2)$ decouple automatically. In the ``two mass dominance''  
scheme, only the pair $(\theta_{12}, \delta_1)$ does. Thus only the exact 
number of physical parameters, according to Table 2, remains both in the 
minimal and next-to-minimal schemes. Notice that it is also important to 
distribute the phases so that they decouple, together with the angles, 
when they should.  

\subsection{Sensitivity reach of the {\em neutrino factory} for four 
neutrino species}

We concentrate now on the sensitivity to the different angles in
  the ``one mass scale'' approximation, discussed at the beginning of 
this section.
Four rotation angles ($\theta_{13}$, $\theta_{14}$, $\theta_{23}$ and 
$\theta_{24}$) and one complex phase ($\delta_3$) remain. The two rotation 
angles that have become unphysical are already tested at solar ($\theta_{12}$ 
in our parametrization) and atmospheric ($\theta_{34}$) neutrino experiments. 
The remaining four can be studied at the {\em neutrino factory} with high 
precision, 
due to the rich flavour content of the neutrino beam. Notice that the number 
of measurable flavour transitions is enough to constraint them. 
Consider the following channels:
\bea
  \bar{\nu}_e & \to  \bar{\nu}_\mu & \to  \mu^+
                             \qquad (\mu^+ {\rm appearance}) \nn \\
  \nu_\mu & \to  \nu_\mu & \to  \mu^- 
                             \qquad (\mu^- {\rm disappearance}) \nn \\
  \bar{\nu}_e & \to  \bar{\nu}_\tau & \to  \tau^+ 
                             \qquad (\tau^+ {\rm appearance}) \nn \\
  \nu_\mu & \to  \nu_\tau & \to  \tau^- 
                             \qquad (\tau^- {\rm appearance}). 
\eea
Separate the CP-even terms,$\PCPC$, from the CP-odd ones, $\PCPV$, 
 in the following way: 
\be
P(\nu_\alpha \to \nu_\beta) = \PCPC(\nu_\alpha \to \nu_\beta) +
                              \PCPV(\nu_\alpha \to \nu_\beta).
\label{dos}
\ee
The CP-even parts acquire simple forms:
\bea
\label{cpeven1}
\PCPC(\nu_e \to \nu_\mu) &=& 4 c^2_{13} c^2_{24} c^2_{23} s^2_{23} \
\sin^2 \left ( \frac{\Delta m_{23}^2 L}{4 E} \right )  \ , \\
\PCPC(\nu_\mu \to \nu_\mu) & = & 1
 - 4 c^2_{13} c^2_{23} ( s^2_{23} + s^2_{13} c^2_{23} ) \
       \sin^2 \left ( \frac{\Delta m_{23}^2 L}{4 E} \right ) \ ,
\label{cpeven2} \\
\PCPC(\nu_e \to \nu_\tau) & = & 4 c^2_{23} c^2_{24} 
\left [ ( s^2_{13} s^2_{14} s^2_{23} + c^2_{14} c^2_{23} s^2_{24} ) 
\right .  \nn \\
& - & \left .  2 c_{14} s_{14} c_{23} s_{23} s_{13} s_{24} 
\cos \delta_3 \right ] \
\sin^2 \left ( \frac{\Delta m_{23}^2 L}{4 E} \right )  \ ,
\label{cpeven3} \\
\PCPC(\nu_\mu \to \nu_\tau) & = & 4 c^2_{13} c^2_{23} 
\left [ ( s^2_{13} s^2_{14} c^2_{23} + c^2_{14} s^2_{23} s^2_{24} ) 
\right . \nn \\
& + & \left .  2 c_{14} s_{14} c_{23} s_{23} s_{13} s_{24} 
\cos \delta_3 \right ] \
\sin^2 \left ( \frac{\Delta m_{23}^2 L}{4 E} \right ).
\label{cpeven4}
\eea
Notice that the physical phase appears in $\PCPC(\nu_e \to \nu_\tau)$ and 
$\PCPC(\nu_\mu \to \nu_\tau)$ in a pure cosine dependence. Actually, no 
CP-odd observable can be build out of the oscillation 
probabilities in this approximation in spite of the existence of a
physical CP-odd phase in the mixing matrix. 

The existing experimental data impose some constraints on the parameter 
space, although the allowed region is still quite large.
Bugey and Chooz are sensitive to the oscillation $\bar{\nu}_e \to \bar{\nu}_e$,
\be
P(\bar{\nu}_e \to \bar{\nu}_e) = 1
 - 4 c^2_{23} c^2_{24} ( s^2_{24} + s^2_{23} c^2_{24} ) 
       \sin^2 \left ( \frac{\Delta m_{23}^2 L}{4 E} \right ),
\ee
resulting in the bound  
\be
( c_{23}^2 \sin^2 2 \theta_{24} + c^4_{24} \sin^2 2 \theta_{23}  ) \le 0.2 \ ,
\label{boundbugey}
\ee
while Bugey gives a slightly stronger constraint in the larger mass range 
allowed by LSND. In our computations we safely stay within both experimental 
constraints. Notice that in the assumption of small angles,  
this bound forces the mixings $s^2_{23}$ and $s^2_{24}$ to be small 
and leaves more freedom in the mixings of the sterile neutrino: 
$s_{13}^2$ and $s_{14}^2$.

The allowed mass range for the LSND signal of $\nu_e \to \nu_\mu$ transitions 
provides the constraint $10^{-3} \leq c^2_{13} c^2_{24} 
\sin^2 2 \theta_{23} \leq 10^{-2}$, which fits nicely with the Chooz constraint 
to point towards small $s_{23}^2$ values.

We choose to be ``conservative'', or even ``pessimistic'', in order to 
illustrate the potential of the {\em neutrino factory}. In the numerical 
computations below we will make the assumption that all angles crossing 
the large LSND gap, $\theta_{13}$, $\theta_{14}$, $\theta_{23}$ and 
$\theta_{24}$ are small.
  
The large LSND mass difference, $\Delta m_{23}^2 \simeq 1 {\rm eV}^2$, 
calls for a SBL experiment rather than a long baseline one. 
Consider for illustration 
a 1 Ton detector located at 
$\simeq 1$ km distance from the neutrino source. 
We consider a muon beam of $E_\mu = 20$ GeV, resulting in $N_{CC} \simeq 10^7$ 
charged leptons detected, for a beam intensity of $2 \times 10^{20}$ useful 
$\mu^-$ per year. An efficiency of $\epsilon_\mu = 0.5, \epsilon_\tau = 0.35$ 
for $\mu$ and $\tau$ detection respectively, and a background contamination 
at the level of $10^{-5} N_{CC}$ events are included. 

\begin{figure}[t]
\vspace{0.1cm}
\centerline{
\epsfig{figure=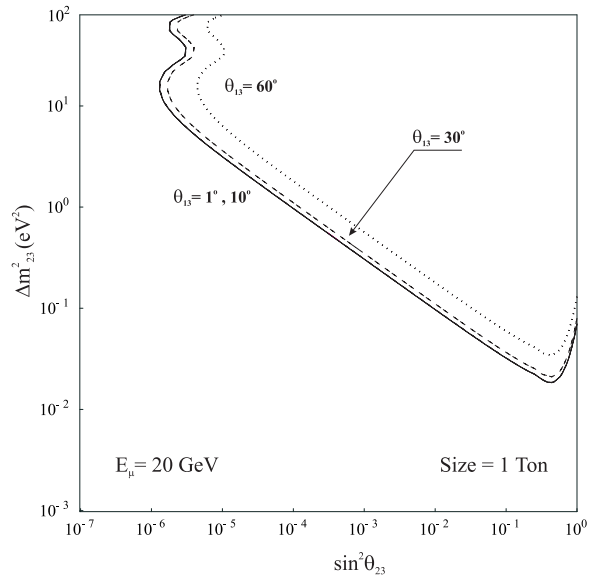,height=7.6cm,angle=0}}
\caption{\it{
Sensitivity reach in the 
$s^2_{23} / \Delta m_{23}^2$ plane at different values of 
$\theta_{13}= 1^\circ, 10^\circ, 30^\circ$ and $60^\circ$
for $\mu^+$ appearance. We consider a 1 Ton detector at $ 1$ km 
from the source and $2 \times 10^{20}$ useful muons/year.}} 
\label{fig:mu23a}
\end{figure}

\begin{itemize}
\item {\bf $\theta_{23}$ and $\theta_{13}$ from $\mu$ channels}
\end{itemize}
The $\mu^+$ appearance channel is particularly sensitive to $\theta_{23}$. 
Fig. \ref{fig:mu23a} shows the sensitivity reach 
in the $s^2_{23} / \Delta m_{23}^2$ plane for different
values of $\theta_{13}$. Inside the LSND allowed region the dependence 
on $\theta_{13}$ is mild: $s_{23}^2$ can reach $10^{-6}$ for $\theta_{13} 
\simeq 1^\circ$ or $6 \times 10^{-6}$ for $\theta_{13} \simeq 60^\circ$. 

\begin{figure}[t]
\vspace{0.1cm}
\begin{tabular}{cc}
\epsfig{figure=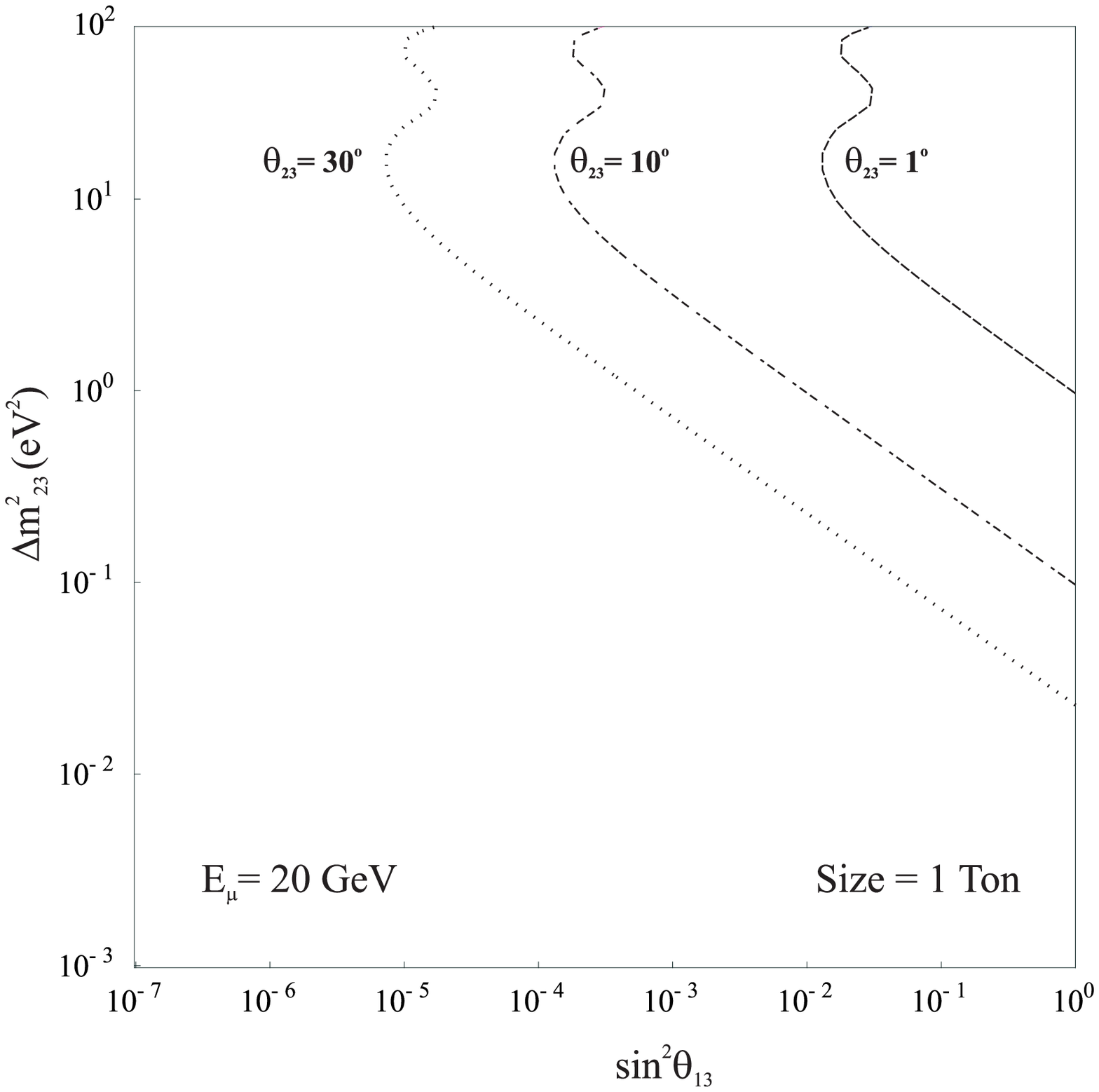,height=7.6cm,angle=0} &
\epsfig{figure=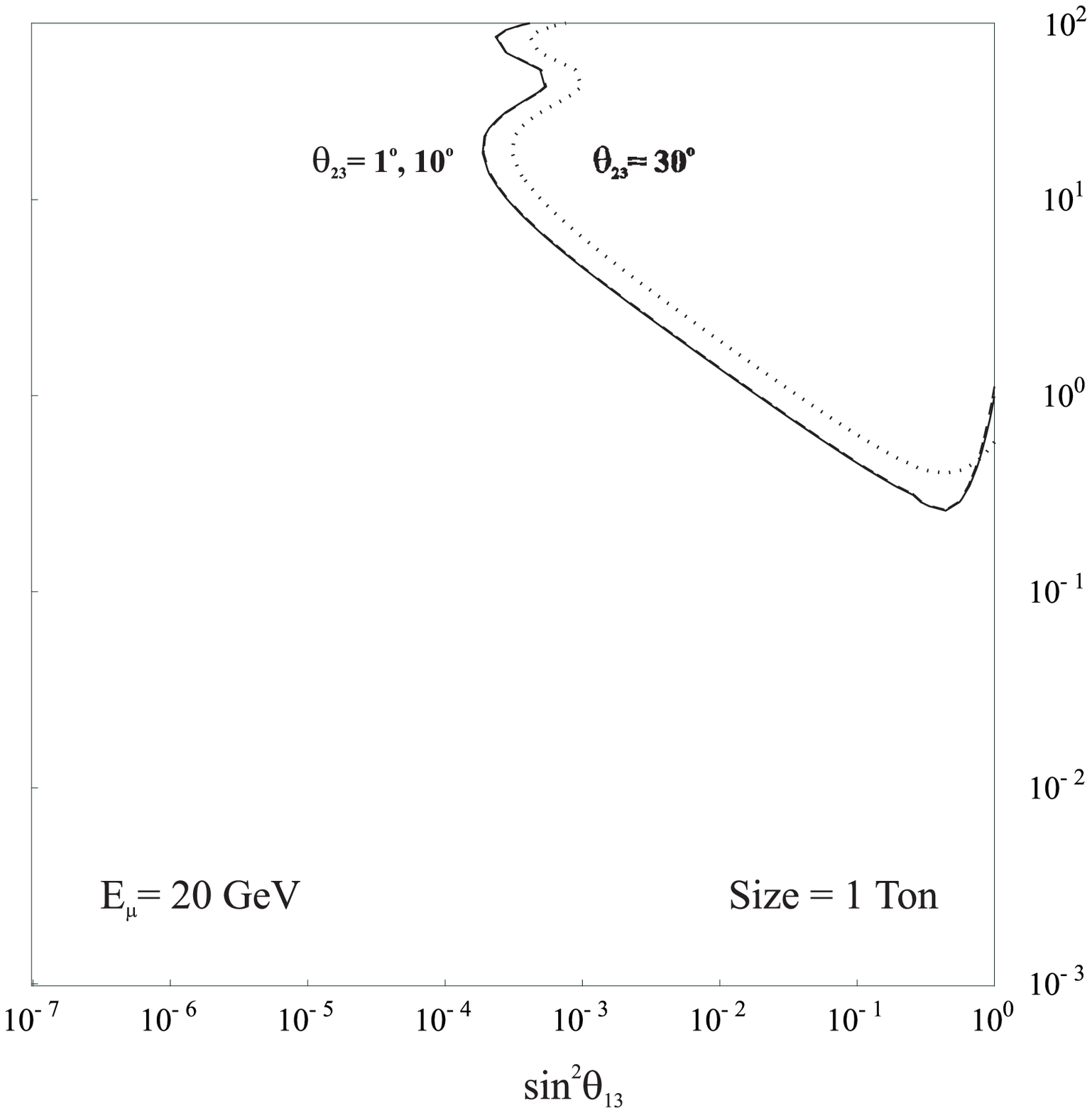,height=7.6cm,angle=0} 
\end{tabular}
\caption{\it{
Sensitivity reach in the 
$s^2_{13} / \Delta m_{23}^2$ plane at different values of 
$\theta_{23}= 1^\circ, 10^\circ$ and $30^\circ$ for $\mu^+$ appearance 
(left) and disappearance (right). We consider a 1 Ton detector at $ 1$ km 
from the source and $2 \times 10^{20}$ useful muons/year.}} 
\label{fig:mu13ad}
\end{figure}

Concerning the sensitivity to $\theta_{13}$, Fig. \ref{fig:mu13ad} (left) 
illustrates the reach from $\mu$ appearance measurements in the 
$s^2_{13} / \Delta m_{23}^2$ plane, for different values of 
$\theta_{23}$. Inside the LSND allowed region, the sensitivity 
to this angle strongly depends on the value
of $\theta_{23}$, with the larger sensitivity attained for large 
values of $\theta_{23}$, a scenario somewhat disfavoured by the LSND 
measurement. For small values of $\theta_{23}$ ($\simeq 1^\circ$), 
the smallest testable value of $s_{13}^2$ is $\sim 10^{-2}$. 
Nevertheless, in this range the muon disappearance channel proves 
quite more sensitive: in Fig.~\ref{fig:mu13ad} (right) the sensitivity 
goes down to values of 
$s_{13}^2$ as small as $10^{-4}$ for $\theta_{23} \simeq  1^\circ$.

\begin{figure}[t]
\vspace{0.1cm}
\centerline{
\epsfig{figure=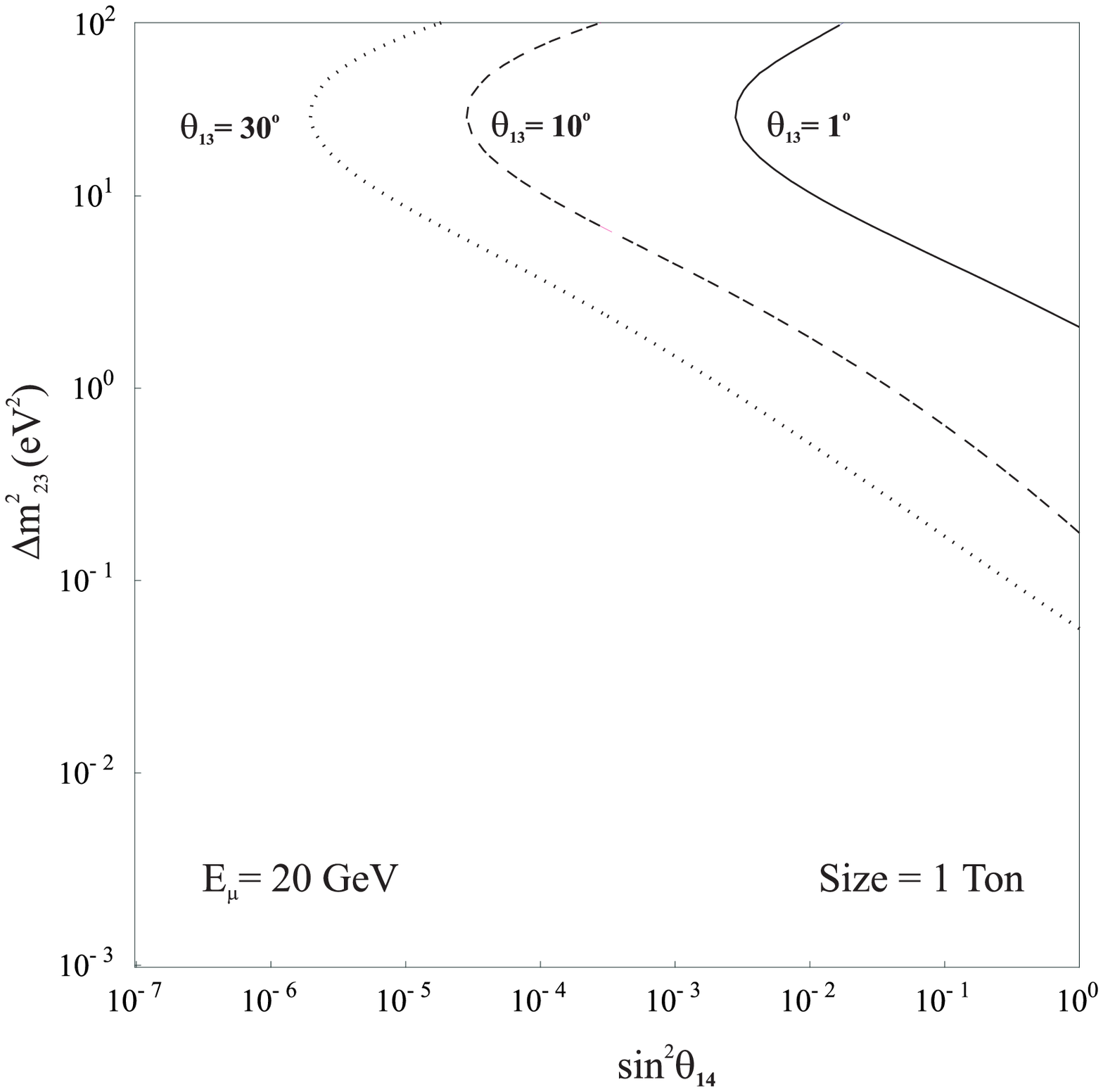,height=7.6cm,angle=0}} 
\caption{\it{
Sensitivity reach in the $s^2_{14}/ \Delta m_{23}^2$ plane at 
different values of $\theta_{13}= 1^\circ, 10^\circ$ and $30^\circ$
for $\tau^-$ appearance. We consider a 1 Ton detector at $1$ km 
from the source and $2 \times 10^{20}$ useful muons/year.}} 
\label{fig:taum14a}
\end{figure}
\begin{figure}[th]
\vspace{0.1cm}
\begin{tabular}{cc}
\epsfig{figure=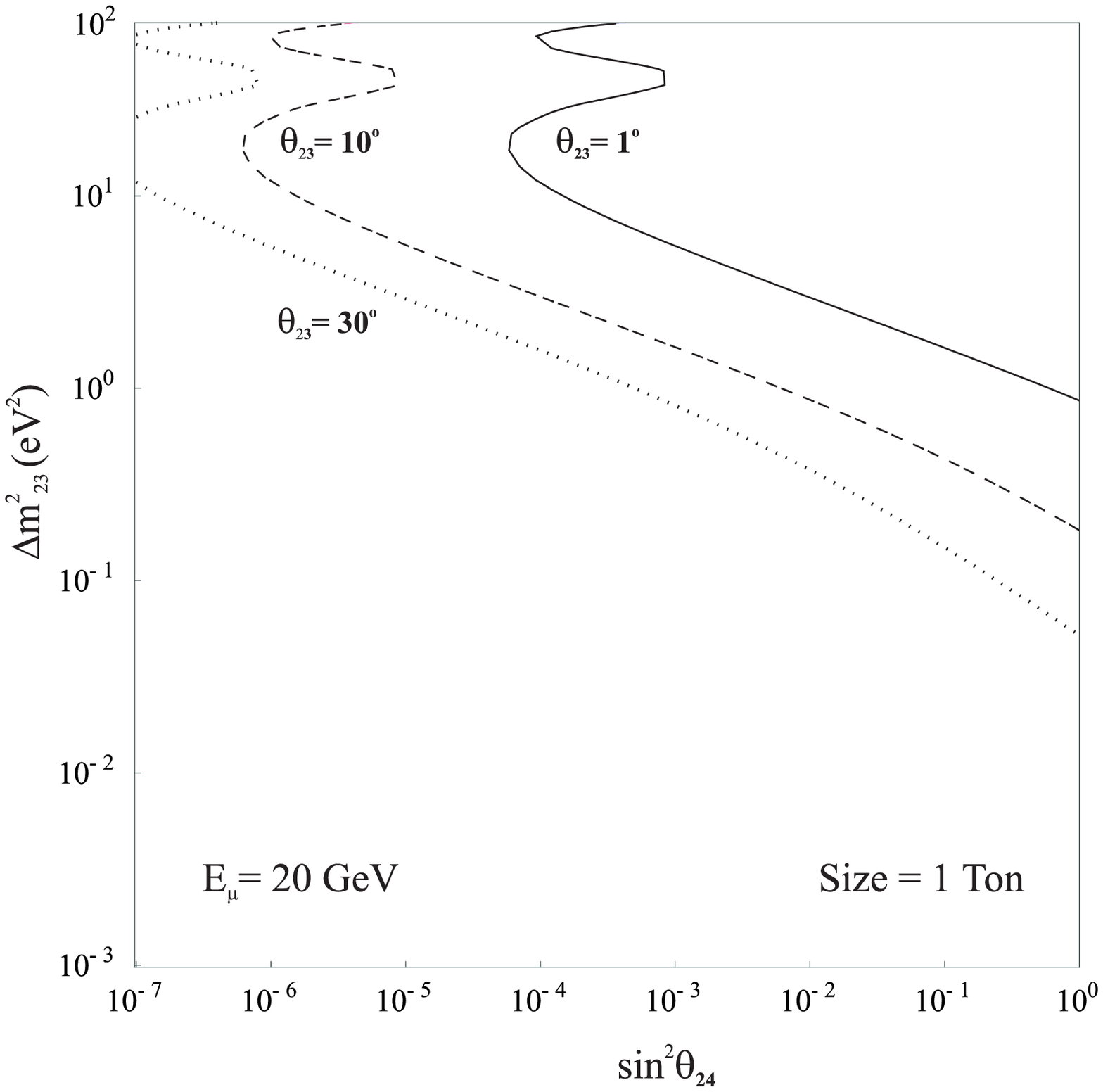,height=7.6cm,angle=0} & 
\epsfig{figure=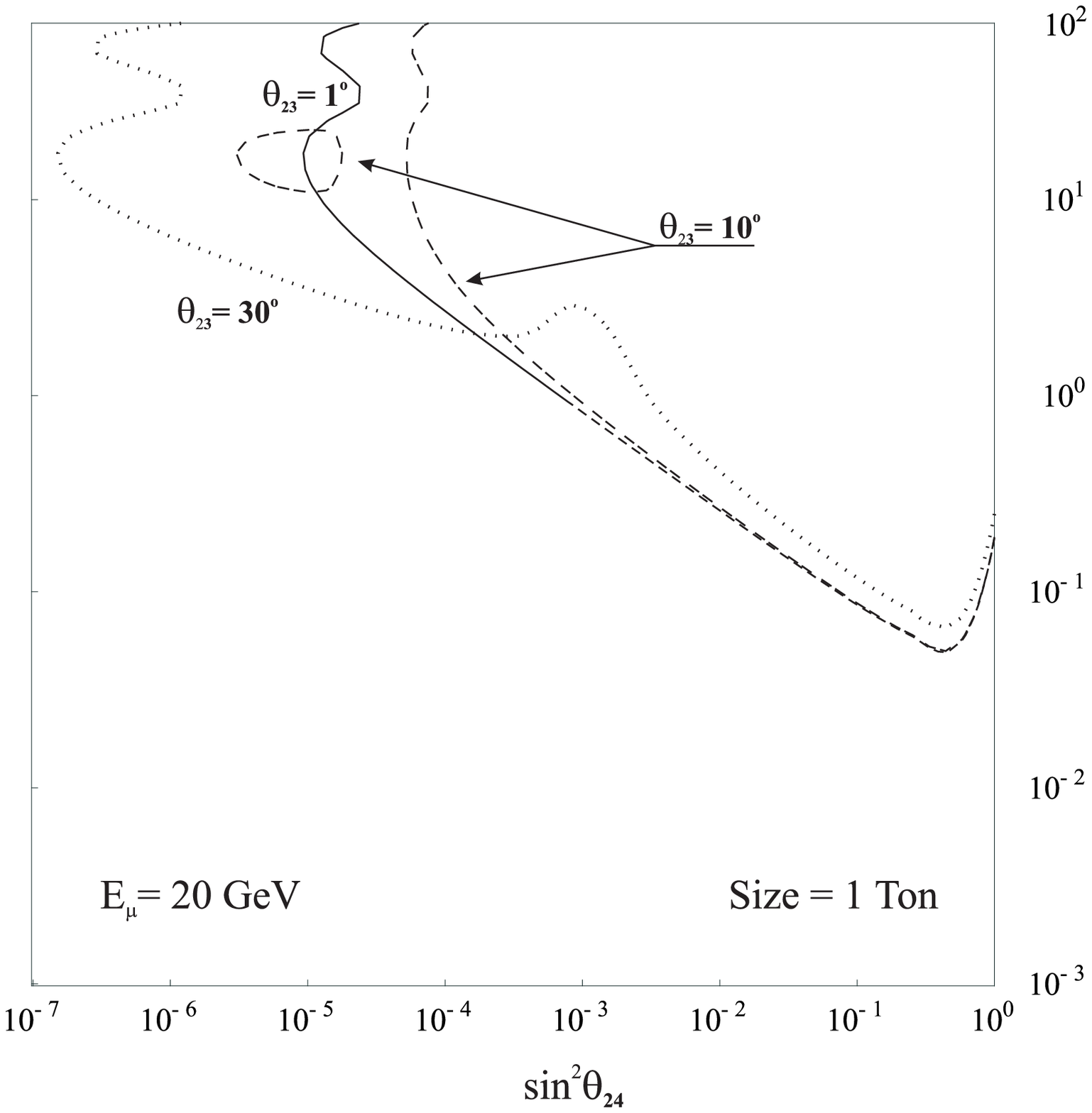,height=7.6cm,angle=0} 
\end{tabular}
\caption{\it{
Sensitivity reach in the $s^2_{24} / \Delta m_{23}^2$ plane at 
different values of $\theta_{23}= 1^\circ, 10^\circ$ and $30^\circ$ for 
$\tau^-$ (left) and $\tau^+$ (right) appearance. We consider a 1 Ton 
detector at $1$ km from the source and $2 \times 10^{20}$ useful muons/year.}} 
\label{fig:tauem24a}
\end{figure}

\begin{itemize}
\item  {\bf $\theta_{14}$ and $\theta_{24}$ from $\tau$ channels}
\end{itemize}

The $\tau^-$ appearance channel is quite sensitive to both $s_{14}^2$ and 
$s_{24}^2$. Fig. \ref{fig:taum14a} illustrates the sensitivity to $s_{14}^2$
as a function of $\theta_{13}$: for about $1^\circ$, sensitivities of 
the order of $10^{-2}$ are attainable, while for $10^\circ$ the 
reach extends to $4 \times 10^{-5}$. For even larger values of $\theta_{13}$ 
it goes down to $10^{-6}$ (we recall that $\theta_{13}$ is not constrained by 
the present experimental bounds).

Fig. \ref{fig:tauem24a} (left) depicts the foreseeable sensitivity reach to 
$s^2_{24}$ as a function of $\theta_{23}$ : for small values of 
$\theta_{23}$ the sensitivity to $s_{24}^2$ goes down to $10^{-6}$. 

In contrast, the $\tau^+$ appearance channel looks less promising.
This is illustrated in Fig. \ref{fig:tauem24a} (right): due 
to the relative negative sign between the two terms in the 
analytic expression for $P(\nu_e \to \nu_\tau)$, eq.~(\ref{cpeven3}),
cancellations for particular values of the angles occur, resulting
in a decreasing sensitivity in specific regions of the parameter space.
For instance, for $\theta_{23} = 10^\circ$, the reach in 
$s^2_{24}$ splits into two separate regions for the LSND allowed range 
$\Delta m^2_{23} \sim 10^1$ eV$^2$.
This sensitivity suppression is absent in the $\tau^-$ channel as the
relative sign between the two terms in $P(\nu_\mu \to \nu_\tau)$, 
eq.~(\ref{cpeven4}), is positive.

The overall conclusion of this analysis is that, in the minimal scheme 
for four-neutrino families, 
a 1 Ton near detector with $\mu$ and $\tau$ charge identification 
is suitable to fully explore the CP-even part of the parameter space. 

\subsection{CP Violation with four light neutrino species}
\label{sect:4CPviolation}
%
As in the standard three-family scenario, in order to reach observable 
CP-odd effects in oscillations it is necessary to have both physical CP-odd 
phases and at least two non-vanishing mass differences. The next-to minimal 
or ``two mass scale dominance'' scheme, described at the beginning of this
section, is thus suitable.

As explained above, the parameter space consists of five angles and two 
CP-odd phases. Expanding the transition probabilities to leading order in 
$\Delta m^2_{atm}$ (i.e. $\Delta m^2_{34}$ in our parametrization), it follows 
that their CP-odd components are\footnote{At this order also sub-leadings in 
the CP-even sector contribute. Although we do no illustrate them, all orders 
in $\Delta m^2_{atm}$ are included in the numerical computations.}:
\bea
\label{cpodd1}
\PCPV(\nu_e \to \nu_e) & = & \PCPV(\nu_\mu \to \nu_\mu)  = 
                             \PCPV(\nu_\tau \to \nu_\tau) = 0 \ , \\
\PCPV(\nu_e \to \nu_\mu) & = & 
     8 c^2_{13} c^2_{23} c_{24} c_{34} s_{24} s_{34} \ 
     \sin (\delta_2 + \delta_3) \
     \left( {{ \Delta m^2_{34} L }\over{4 E_\nu} } \right)  
     \sin^2 \left (\frac{\Delta m_{23}^2 L}{4 E_\nu} \right)  \\
\PCPV(\nu_e \to \nu_\tau) & = & 4 c_{23} c_{24} 
      \Big \{ 
      2 c_{14} s_{14} c_{23} s_{23} s_{13} s_{24} 
      (s^2_{13} s^2_{14} - c^2_{14}) \sin (\delta_2 + \delta_3) \nn \\
& + & 
      c_{14} c_{34} s_{13} s_{14} s_{34} \left[ 
      (s^2_{23} - s^2_{24}) \ \sin \delta_2 + 
      s^2_{23} s^2_{24} \sin(\delta_2 + 2\delta_3) \right] \\
& + & 
      c_{14} c_{24} s_{13} s_{14} s_{23} s_{24} 
      (c^2_{34} - s^2_{34}) \ \sin \delta_3  \Big \}
      \left( {{ \Delta m^2_{34} L }\over{4 E_\nu} } \right)   
      \sin^2 \left( \frac{\Delta m_{23}^2 L}{4 E_\nu} \right) \nn \\
\PCPV(\nu_\mu \to \nu_\tau) & = & 
      8 c^2_{13} c^2_{23} c_{24} c_{34} s_{34} \left [ 
      c_{14} c_{23} s_{13} s_{14} \ \sin \delta_2 \ + 
      c^2_{14} s_{23} s_{24} \ \sin (\delta_2 + \delta_3) \right ] \times \nn \\ 
&   & \qquad \left( {{ \Delta m^2_{34} L }\over{4 E} } \right )  
      \sin^2 \left( \frac{\Delta m_{23}^2 L}{4 E} \right ).
\label{cpodd2}
\eea

Two distinct phases appear, $\delta_2$ and $\delta_3$, 
in a typical sinusoidal dependence which is the trademark of CP-violation 
and ensures different transition rates for neutrinos and anti neutrinos.

CP-odd effects are observable in ``appearance'' channels, while 
``disappearance'' ones are only sensitive to the CP-even part. The latter 
is mandated by CPT \cite{bernabeu}. In contrast with the three-neutrino
case, the solar suppression (see \cite{dgh}) is now replaced by a less 
severe atmospheric suppression. CP-violating effects are then expected to 
be one or two orders of magnitude larger that in the standard case, and 
independent of the solar parameters. 

Staying in the ``conservative'' assumption of small $\theta_{13},\theta_{14}, 
\theta_{23}, \theta_{24}$, we compare two democratic scenarios, in which 
all these angles are taken to be small and of the same order:
\begin{enumerate}
\item Set 1: $\theta_{34} = 45^\circ$, $\theta_{ij} = 5^\circ$ 
      and $\Delta m^2_{atm} = 2.8 \times 10^{-3}$ eV$^2$
      for $\Delta m^2_{LSND} = 0.3$ eV$^2$;
\item Set 2: $\theta_{34} = 45^\circ$, $\theta_{ij} = 2^\circ$ and 
      $\Delta m^2_{atm} = 2.8 \times 10^{-3}$ eV$^2$ for 
      $\Delta m^2_{LSND} = 1$ eV$^2$.
\end{enumerate}
The value chosen for $\Delta m^2_{atm} $ is the central 
one of the most recent SuperK analysis \cite{Superka}. For illustration we 
consider in what follows a 1 kTon detector located at 
$O(10)$ km distance from the neutrino source. In the figures below, 
the exact formulae for the probabilities have been used.

The easiest way to measure CP-violation in oscillation is to built a 
CP-asymmetry or a T-asymmetry:
\be
A_{\alpha \beta}^{CP} \equiv \frac{
P(\nu_\alpha\raw \nu_\beta)-P(\bar{\nu}_\alpha \raw \bar{\nu}_\beta)}
{P(\nu_\alpha \raw \nu_\beta)+P(\bar{\nu}_\alpha \raw \bar{\nu}_\beta)} \ ,
\label{CPodd}
\ee
\be 
A_{\alpha \beta}^{T} \equiv \frac{
P(\nu_\alpha\raw \nu_\beta) - P(\nu_\beta \raw \nu_\alpha)}
{P(\nu_\alpha \raw \nu_\beta) + P(\nu_\beta \raw \nu_\alpha)} \ .
\label{Todd} 
\ee
$A_{\alpha \beta}^{CP}$ and $A_{\alpha \beta}^{T}$ are theoretically 
equivalent in vacuum due to $CPT$, and matter effects are negligible
at the short distances under consideration. Their extraction from
 data at a  {\em neutrino factory}  is quite different, though.
Consider, as an example, the $(e\mu)$-channel. The CP-asymmetry, 
$A_{e \mu}^{CP}$, would be measured by first
extracting $P(\nu_\mu\raw \nu_e)$ from the produced (wrong-sign) $\mu^-$s 
in a beam from $\mu^+$ decay and $P(\bar\nu_e\raw \bar\nu_\mu)$ from the 
charge conjugate beam and process. Notice that even if the fluxes are very 
well known, this requires a good knowledge of the cross section ratio 
$\sigma(\bar\nu_\mu\to\mu^+)/\sigma(\nu_\mu\to\mu^-)$. Conversely, the 
measurement of the T-asymmetry, $A_{e \mu}^{T}$, requires  
to consider $P(\nu_\mu \raw \nu_e)$ and 
thus a good $e$ charge identification, that seems harder to achieve.
In the following we will deal only with CP-asymmetries.

A central question on the observability of CP-violation is that of statistics.
We do not exploit here the 
explicit $E_\nu$ dependence of the CP-odd effect, and we consider the 
neutrino-energy integrated 
quantity: 
\begin{equation}
{\bar A}^{CP}_{e\mu} (\delta) = \frac{\{{N[\mu^-]}/{N_o[e^-]}\}_{+} 
-\{N[\mu^+]/N_o[e^+]\}_{-}}{\{N[\mu^-]/N_o[e^-]\}_{+} 
+\{N[\mu^+]/N_o[e^+]\}_{-}}\; ,
\label{intasy}
\end{equation}
where the sign of the decaying muons is indicated by a subindex,
$N[\mu^+]$ $(N[\mu^-])$ are the measured number of wrong-sign muons, and 
$N_o[e^+]$ $(N_o[e^-])$ are the expected number of $\bar{\nu}_e (\nu_e)$ 
charged current interactions in the absence of oscillations.
In order to quantify the significance of the signal, we compare the 
value of the integrated asymmetry with its error, in which we include the 
statistical error and a conservative background estimate at
the level of $10^{-5}$.

The size of the CP-asymmetries is very different for $\mu$ channels and 
$\tau$ channels. For instance for Set 2, they turn out to be small 
in $\nu_e-\nu_\mu$ oscillations, ranging from the per mil level to a few percent. 
In contrast, in $\nu_\mu-\nu_\tau$ oscillations they attain much larger values 
of about $50\%-90\%$. This means that their hypothetical measurement should be 
rather insensitive to systematic effects, and other conventional neutrino beams 
from pion and kaon decay could be appropriate for their study.

%
\begin{figure}[t]
\vspace{0.1cm}
\begin{tabular}{cc}
\epsfig{figure=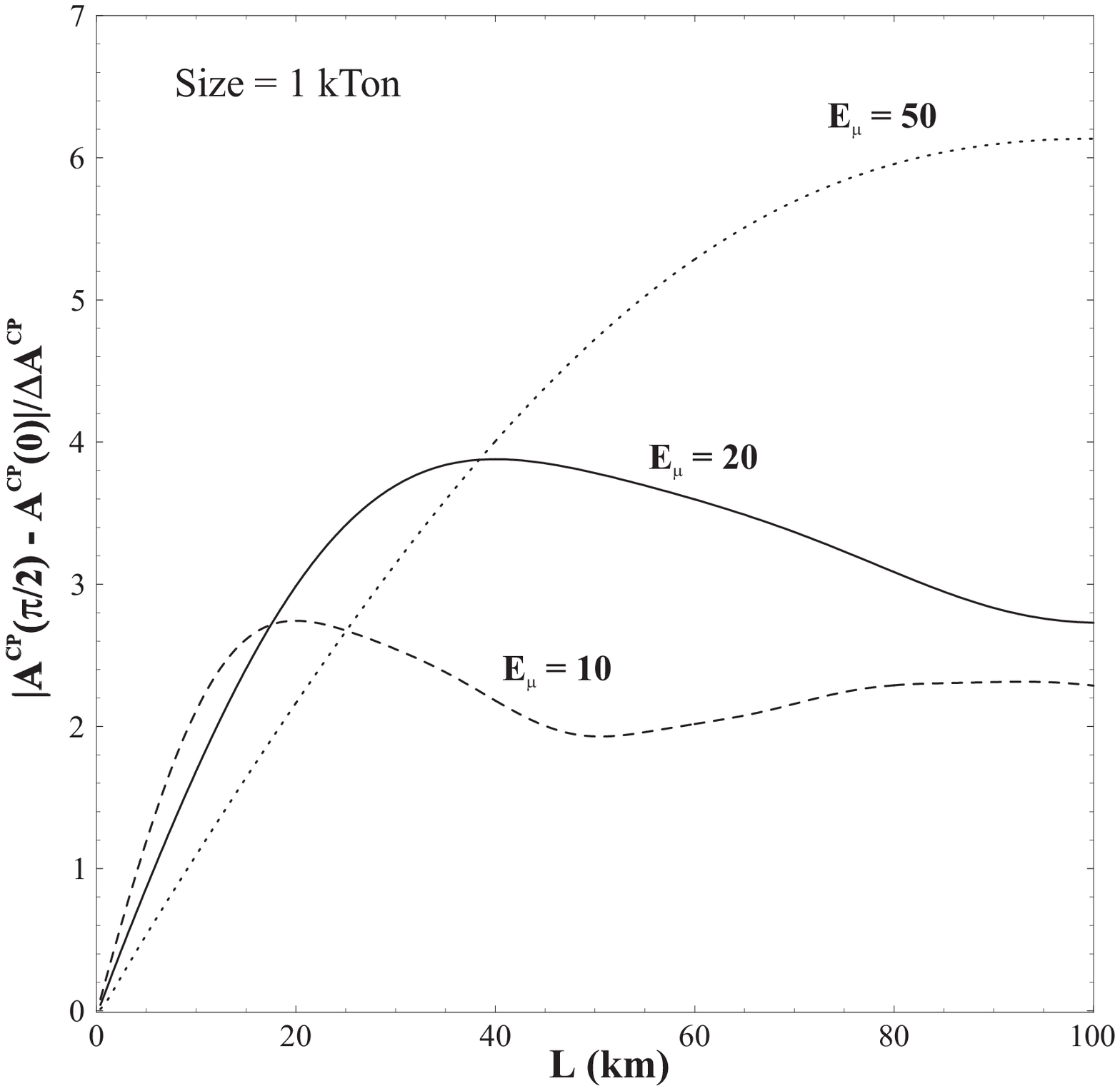,height=7.5cm,angle=0} & 
\epsfig{figure=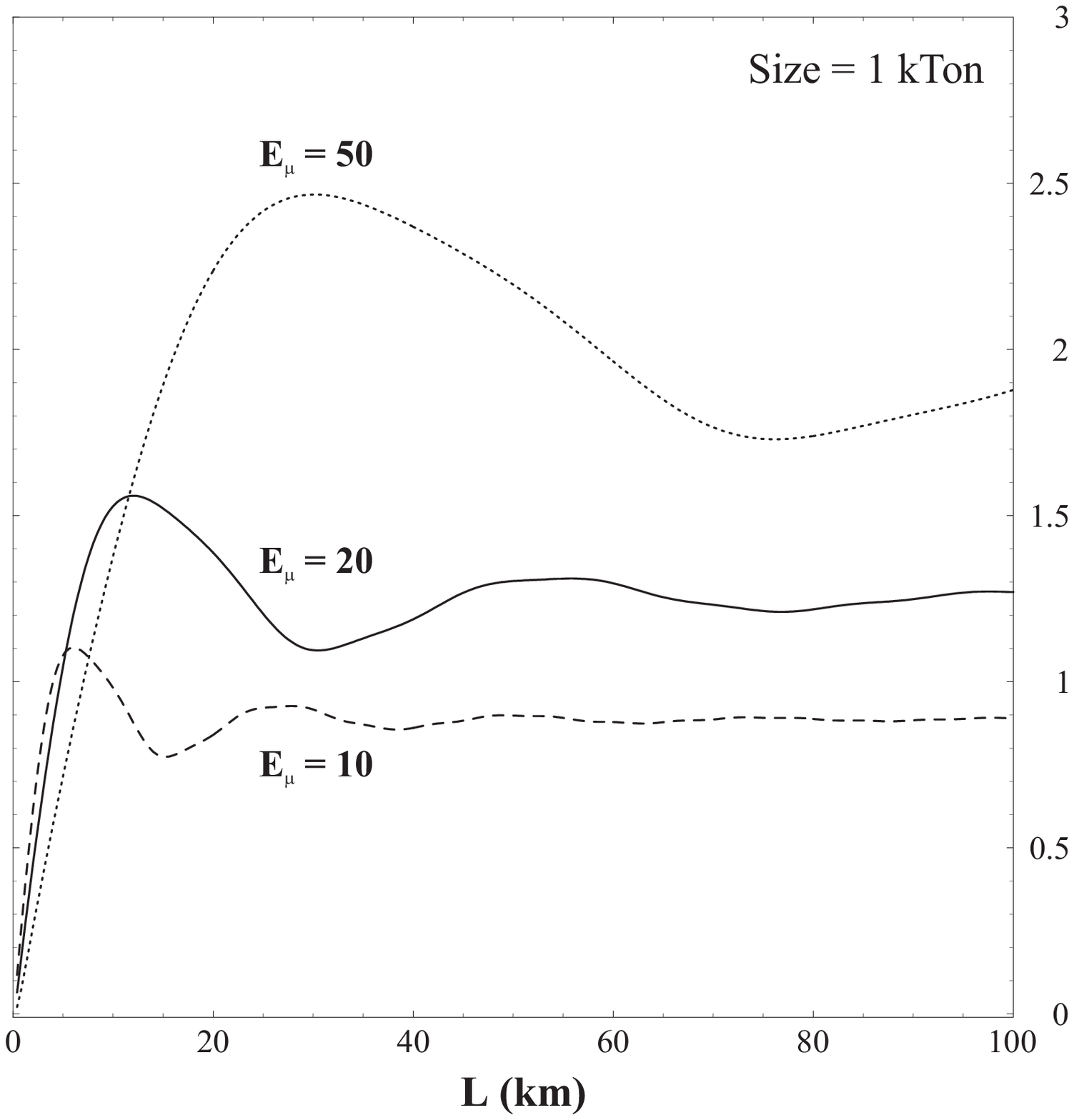,height=7.6cm,angle=0} 
\end{tabular}
\caption{\it{
Signal over statistical uncertainty for CP violation, in the 
$\nu_e \to \nu_\mu$ channel, for the two sets of parameters described in the 
text (Set 1 on the left and Set 2 on the right). We consider a $1$ kTon 
detector and $2 \times 10^{20}$ useful muons/year.}} 
\label{CPetfig12}
\end{figure}
%
%
\begin{itemize}
\item {\bf $\mu$ appearance channels}
\end{itemize}
Fig.~\ref{CPetfig12} shows the signal over noise ratio for the integrated 
CP asymmetry, eqs. (\ref{intasy}), in the wrong 
sign muon channel, that is $\nu_e \to 
\nu_\mu$ versus $\bar\nu_e \to \bar\nu_\mu$ oscillations, as a function of 
the distance. Matter effects, although negligible, have been included.
For the scenario and distances discussed here, the scaling laws are analogous 
to those derived for three neutrino species in vacuum, \cite{dgh,noi,Donini:1999jc}, 
that is
\be
\frac{A_{e \mu}^{CP}}{\Delta A_{e \mu}^{CP}} \propto \sqrt{E_\nu}  
\left| \sin \left ( \frac{\Delta m^2_{34} \, L}{4 E_\nu} \right ) \right| .
\label{scaling2}
\ee
The maxima of the curves move towards larger distances when the energy of the 
muon beam is increased, or the assumed LSND mass difference is decreased. 
Moreover, increasing the energy enhances the significance of the effect at the 
maximum as expected. 
At $E_\mu = 50$ GeV, 6 standard deviation (sd) signals are attainable at 
around 100 km for the values in Set 1, and just 2.5 sd at 30 km for Set 2, 
levelling off at larger distances and finally diminishing when $E_\nu/L$ 
approaches the atmospheric range.

%
\begin{figure}[t]
\vspace{0.1cm}
\begin{tabular}{cc}
\epsfig{figure=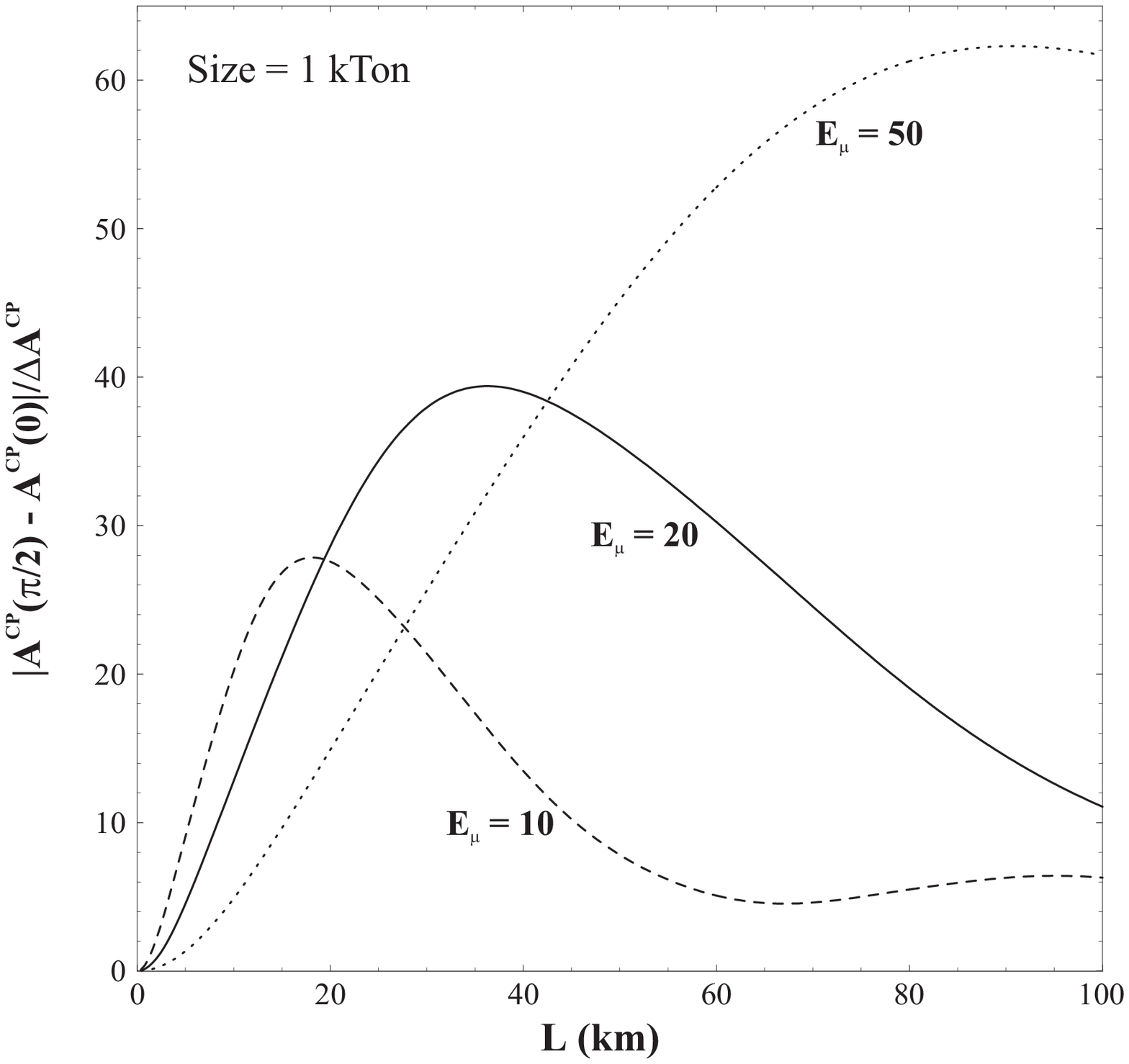,height=7.5cm,angle=0} & 
\epsfig{figure=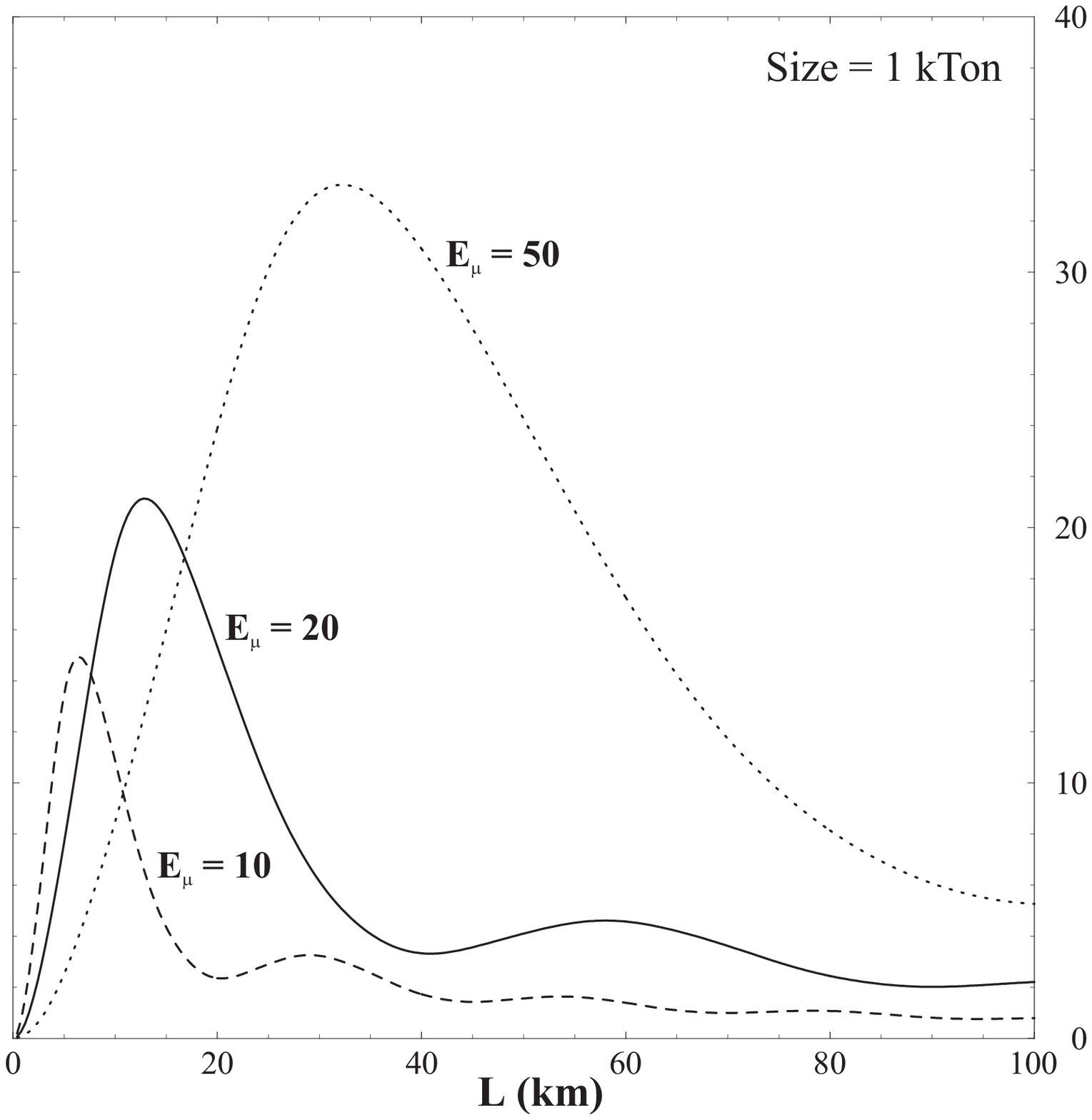,height=7.5cm,angle=0} 
\end{tabular}
\caption{\it{
Signal over statistical uncertainty for CP violation,
in the $\nu_\mu \to \nu_\tau$ channel, for the two sets of parameters described 
in the text (Set 1 on the left and Set 2 on the right). We consider a $1$ kTon 
detector and $2 \times 10^{20}$ useful muons/year.}} 
\label{CPmtfig12}
\end{figure}
%
\begin{itemize}
\item {\bf $\tau$ appearance channels}
\end{itemize}
In Fig.~\ref{CPmtfig12} we show the signal over noise ratio in $\nu_\mu 
\to \nu_\tau$ versus $\bar\nu_\mu \to \bar\nu_\tau$ oscillations as a 
function of the distance. The experimental asymmetry and the corresponding  
scaling law are obtained from eq.~(\ref{intasy}) and eq.~(\ref{scaling2}), 
with the obvious replacements $e \to \mu$ and $\mu \to\tau$.
A larger enhancement takes place in this channel as compared to the 
$\nu_e \to \nu_\mu$ one: over 60 sd for Set 1 and 33 sd for Set 2 are a 
priori attainable. These larger factors follow from the fact that the CP-even 
transition probability $P_{CP}(\nu_\mu \nu_\tau)$ is smaller than 
$P_{CP}(\nu_e \nu_\mu)$, due to a  stronger suppression in small mixing angles.
Notice that the opposite happens in the 3-species case. 
Bilenky {\em et al.} \cite{bilenky} had previously concluded that 
the $\tau$ channel was best for CP-studies in the four species scenario. 
Their argument relied, though, on the fact that the parameter space involved 
in $\nu_\tau$ oscillations  is experimentally less constrained than the 
$\nu_{\mu}$ one, a freedom we have not used here, staying within the more 
natural assumption that all the angles in the next-to-minimal scheme,
except the atmospheric one, $\theta_{34}$, are small.

The results in the $\nu_e \to \nu_\tau$ channels are almost identical   
to the $\nu_e \to \nu_\mu$ ones, not deserving a separate discussion.

\begin{figure}[t]
\vspace{0.1cm}
\begin{tabular}{cc}
\epsfig{figure=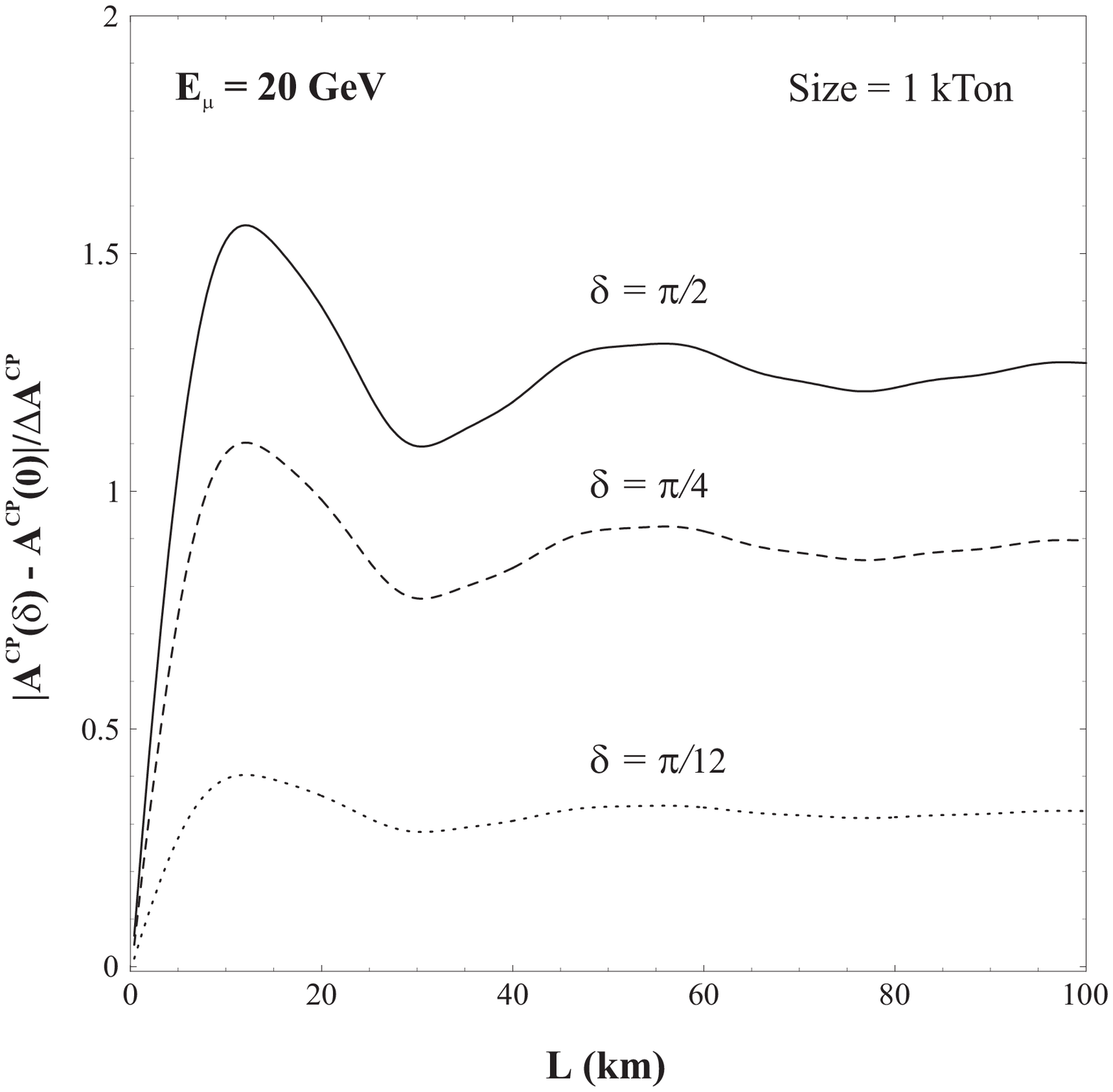,height=7.6cm,angle=0} & 
\epsfig{figure=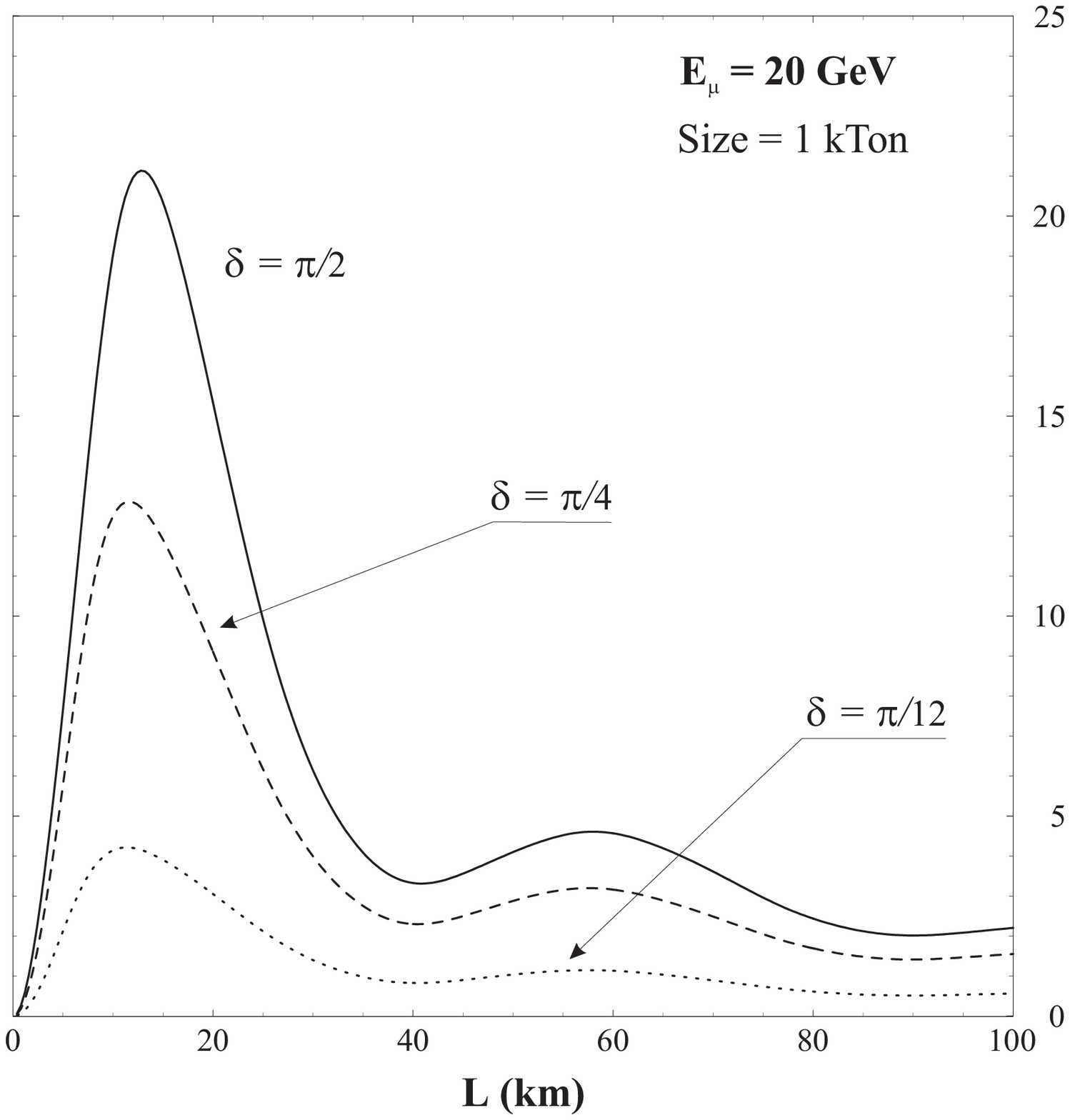,height=7.6cm,angle=0} 
\end{tabular}
\caption{\it
CP violation asymmetry in the $\nu_e \to \nu_\mu$ (left) and 
$\nu_\mu \to \nu_\tau$ (right) channel for $E_\mu=20$ GeV, angles and mass 
differences as in Set 2 and for different choice of the CP phases: 
$\delta_1=\delta_2=\delta_3=\pi/2$ (full line), $\pi/4$ (dashed line) and 
$\pi/12$ (dotted line). We consider a $1$ kTon detector from the 
source of a $2 \times 10^{20}$ muon/year beam.} 
\label{CPphfig}
\end{figure}

The phase dependence is shown in Fig.~\ref{CPphfig}, with the expected 
depletion of the signal for small CP phases. For small values of the phases, 
i.e. $\delta_1=\delta_2=\delta_3 = 15^\circ$, the significance drops to 
the $1 \sigma$ level. 

%
%
%
%


\section{Conclusions}

The ensemble of solar, atmospheric and LSND neutrino data, are analysed in a 
three active plus one sterile scenario. A {\em neutrino factory} from muon 
storage beams has much higher precision and discovery potential than any 
other planned facility. The reach of SBL experiments is extremely large. 
We have derived one and two mass scale dominance approximations, appropriate 
for CP-even and CP-odd observables, respectively. The number of useful 
observables is sufficient to determine or very significantly constrain all 
the mixing parameters of a four-generation mixing scheme. CP violation may 
be easily at reach, specially through ``$\tau$ appearance'' signals. In these 
channels the CP-asymmetries are so large that even neutrino beams from 
conventional pion and kaon decays may be sufficient for their detection.     
Matter effects are irrelevant in the energy integrated CP-odd observables 
for such a short baseline, making the measure of CP-violation extremely 
clean in this four neutrino scenario.

\section{Acknowledgements}
We acknowledge useful conversations with: B. Autin, A. De R\'ujula, 
F. Dydak, J. G\'omez-Cadenas, M.C. Gonzalez-Garcia, O. Mena, S. Petcov 
and A. Romanino. 
The work of A. D., M. B. G., and S. R. 
was partially supported 
by CICYT project AEN/97/1678. A. Donini acknowledges the I.N.F.N. 
for financial support. S. Rigolin acknowledges the European Union for 
financial support through contract ERBFMBICT972474.



\end{document}